\documentclass[12pt,a4paper]{article}
\usepackage{graphicx,cite}
\newcommand{\Frac}[2]{\frac{\displaystyle #1}{\displaystyle #2}}

\newcommand{\nn}{\nonumber}

\newcommand\lsim{\mathrel{\rlap{\lower4pt\hbox{\hskip1pt$\sim$}}
    \raise1pt\hbox{$<$}}}
\newcommand\gsim{\mathrel{\rlap{\lower4pt\hbox{\hskip1pt$\sim$}}
    \raise1pt\hbox{$>$}}}

\newcommand{\VAP}{\langle V \! A P\rangle}
\newcommand{\order}{{\cal O}}

\newcommand{\bea}{\begin{eqnarray}}
\newcommand{\eea}{\end{eqnarray}}
\newcommand{\beq}{\begin{equation}}
\newcommand{\eeq}{\end{equation}}
\newcommand{\nl}{\nonumber\\}

\newcommand{\cO}{{\cal O}}

\newcommand{\mrm}{\mathrm} 
\begin{document}

\begin{titlepage}

\begin{flushright}
{\small\sf  MAP-295 \\ IFIC/04-12\\FTUV/04-0401\\UWThPh-2004-6}
\end{flushright}

\vspace*{1.5cm} 
\begin{center}
{\Large\bf $\VAP$ Green Function in the Resonance Region$^*$}
\\[20mm]

{\normalsize\bf \sc V. Cirigliano$^{1}$, G. Ecker$^2$, M.~Eidem\"uller$^{3}$, 
A. Pich$^{3}$ and J. Portol\'es$^{3}$  }\\
 
\vspace{1cm} 
${}^{1}$ Department of Physics, California Institute of Technology\\ 
Pasadena, California 91125, USA\\[10pt]
${}^{2)}$ Institut f\"ur Theoretische Physik, Universit\"at 
Wien\\ Boltzmanngasse 5, A-1090 Vienna, Austria \\[10pt]
${}^{3)}$ Departament de F\'{\i}sica Te\`orica, IFIC, CSIC --- 
Universitat de Val\`encia \\ 
Edifici d'Instituts de Paterna, Apt. Correus 22085, E-46071 
Val\`encia, Spain \\
\end{center}

\vfill

\begin{abstract}
\noindent
We analyse the $\VAP$ three-point function of vector, axial-vector and
pseudoscalar currents. In the spirit of large $N_C$, a resonance
dominated Green function is confronted with the leading high-energy
behaviour from the operator product expansion. The matching is shown
to be fully compatible with a chiral resonance Lagrangian and it
allows to determine some of the chiral low-energy constants of
$\cO(p^6)$. 
\end{abstract}

\vfill


\noindent 
*~Work supported in part by HPRN-CT2002-00311 (EURIDICE) and by
Acciones Integradas, HU2002-0044 (MCYT, Spain), Project No. 19/2003 
(Austria).
\end{titlepage}



\addtocounter{page}{1} 


\paragraph{1.}

Following the work of Moussallam \cite{Moussallam:1997xx} and 
of Knecht and Nyffeler \cite{Knecht:2001xc}, we
reanalyse the three-point function of vector, axial-vector and
pseudoscalar currents. The procedure of matching between low and high
energies is especially transparent for Green functions like the
$\VAP$ correlator that are order parameters of chiral symmetry
breaking. 
Working in the chiral limit and at leading order in the $1/N_C$ expansion, 
the matching is 
performed by saturating the operator product expansion (OPE) with a 
certain number of resonance
multiplets in accordance with the Minimal Hadronic Ansatz \cite{MHA}.
An alternative way consists in using directly a chiral
resonance Lagrangian to perform the matching, getting thereby also
information on some of the resonance couplings of such a Lagrangian.

In Ref.~\cite{Knecht:2001xc} it was claimed that, 
unlike the situation at $\cO(p^4)$ \cite{Ecker:1989te,Ecker:1989yg}, a minimal
chiral resonance Lagrangian using the Proca formalism for spin-1 fields 
\cite{Prades:1994ys} is unable to recover the asymptotic behaviour of 
the Green functions considered, in particular the $\VAP$ correlator.
This claim should be contrasted with the results of 
Ref.~\cite{Ruiz-Femenia:2003hm} for the $\langle VVP \rangle$
correlator where a chiral resonance Lagrangian, using the
antisymmetric tensor formalism of spin-1 fields, was found to be 
consistent with the leading asymptotic behaviour of QCD.

An important bonus of the matching procedure is 
the prediction of resonance contributions to couplings of 
the Chiral Perturbation Theory (CHPT) Lagrangian at ${\cal O}(p^6)$ 
\cite{Fearing:1996ga,Bijnens:1999sh,Bijnens:2001bb,Ebertshauser:2001nj}. 
With 90 such low-energy constants (LECs) in the even-intrinsic parity 
Lagrangian for three flavours, such predictions are indispensable at
present to do
phenomenology to next-to-next-to-leading order in the low-energy
expansion. Compatibility between the constraints from QCD at high
energies and a suitable chiral resonance Lagrangian, as already 
established at $\cO(p^4)$ \cite{Ecker:1989te,Ecker:1989yg}, would 
allow for a consistent Lagrangian treatment of resonance
  contributions at least up to $O(p^6)$.
With this motivation in mind, we first show that a
chiral resonance Lagrangian with an appropriate set of
resonance fields is fully consistent with QCD constraints for the 
$\VAP$ Green function to be specified below. We then go on to 
determine six of the LECs of
$\cO(p^6)$  from $V, A$ and $P$ resonance contributions.   
 
The approach employed in this letter should be seen as an
  approximation to large-$N_C$ QCD. The approximation consists in the
  choice of a hadronic ansatz and in a set of QCD short-distance
  constraints to be satisfied. We adopt the following guidelines:
\begin{itemize} 
\item {\it Hadronic ansatz}: We include all lowest-lying resonance
  multiplets that can contribute to the given Green function. This
  choice is motivated by the well-founded assumption that the
  low-lying hadronic spectrum determines the chiral LECs that govern
  the low-energy behaviour of the Green function.  

\item {\it Short-distance constraints}: Working with a finite number
  of resonance multiplets, conflicts may arise between different types
  of asymptotic constraints \cite{Bijnens:2003rc}. We assign the
  highest priority to the constraints dictated by the OPE to leading
  order in inverse powers of large momenta. We then consider the
  constraints implied by the high-momentum behaviour of all hadronic
  form factors (as predicted by the quark counting rules
  \cite{Lepage:1979zb,Lepage:1980fj,Brodsky:1981rp}) with on-shell
  Goldstone modes and photons. We do not consider form factors with
  external resonance states.

\end{itemize}

\paragraph{2.}  

The $\VAP$ three-point function in momentum space is defined as
\begin{eqnarray}
\label{VAPdef}
(\Pi_{V\!AP})_{\mu\nu}^{abc}(p,q) = \int d^4x \int d^4y \ e^{i(p \cdot x + 
q \cdot y)} \langle 0 \vert T \{ V_\mu^a(x) A_\nu^b(y) P^c(0) \} \vert 0
\rangle \,,
\end{eqnarray}
with $SU(3)$ octet vector, axial-vector and pseudoscalar currents 
\begin{eqnarray}
\label{currents}
V_\mu^a = \bar \psi \gamma_\mu \displaystyle\frac{\lambda^a}{2} \psi \,, 
& \quad
A_\mu^a = \bar \psi \gamma_\mu \gamma_5
\displaystyle\frac{\lambda^a}{2}  \psi \,, 
& \quad
P^a = \bar \psi i \gamma_5 \displaystyle\frac{\lambda^a}{2} \psi \,. 
\end{eqnarray}
It satisfies the chiral Ward identities \cite{Moussallam:1997xx,Gasser:1984yg}
\begin{eqnarray}
\label{WardIdentities}
p^{\mu}(\Pi_{V\!AP})_{\mu\nu}^{abc}(p,q) &=& 
\langle{\overline\psi}\psi\rangle_0 f^{abc}\,\Bigg[\frac{q_{\nu}}{q^2}\,-\,
\frac{(p+q)_{\nu}}{(p+q)^2}\Bigg]\, ,\nn\\
q^{\nu}(\Pi_{V\!AP})_{\mu\nu}^{abc}(p,q) &=&
\langle{\overline\psi}\psi\rangle_0 f^{abc}\,\frac{(p+q)_{\mu}}{(p+q)^2}\, ,
\end{eqnarray}
where $\langle{\overline\psi}\psi\rangle_0$ denotes the
quark condensate in the chiral limit.  The general solution
of these Ward identities, taking into account the QCD symmetries
$SU(3)_V$, parity and time reversal, is 
\cite{Moussallam:1997xx,Knecht:2001xc}
\begin{eqnarray}
\label{GeneralSolution}
(\Pi_{V\!AP})_{\mu\nu}^{abc}(p,q) & = & 
  f^{abc}\,\Bigg\{ \langle{\overline\psi}\psi\rangle_0 \left[
\frac{(p+2q)_\mu q_\nu}{q^2 (p+q)^2} - \frac{g_{\mu\nu}}{(p+q)^2}\right] 
 \nn\\
&& + P_{\mu\nu}(p,q) {\cal F}(p^2,q^2,(p+q)^2) 
+ Q_{\mu\nu}(p,q) {\cal G}(p^2,q^2,(p+q)^2) \Bigg\} \, .
\end{eqnarray}
The transverse tensors $P_{\mu\nu}$ and 
$Q_{\mu\nu}$ are defined as
\begin{eqnarray}
\label{PQdef}
P_{\mu\nu}(p,q) &=& q_\mu p_\nu - (p \cdot q) g_{\mu\nu} \, , \nn\\ 
Q_{\mu\nu}(p,q) &=& p^2 q_\mu q_\nu + q^2 p_\mu p_\nu - (p \cdot q)
p_\mu q_\nu - p^2 q^2 g_{\mu\nu} \, .  
\end{eqnarray}

The behaviour of the invariant functions ${\cal F}$ and ${\cal G}$ 
at small momentum transfers is governed by the contributions from 
Goldstone boson intermediate states. As one-particle
exchange dominates in the limit $N_C \to \infty$, we only need to
keep the corresponding Goldstone boson poles and the polynomial terms
involving the LECs. In the basis of Ref. \cite{Bijnens:1999sh} 
for the LECs of ${\cal O}(p^6)$ one finds \cite{Knecht:2001xc}
($F$ is the pion decay constant in the chiral limit)
\begin{eqnarray}
\label{ChPTSolution}
\! 
{\cal F}^{\mrm{CHPT}}(p^2,q^2,(p+q)^2) & = &  \frac{4 
\langle{\overline\psi}\psi\rangle_0}{F^2 (p+q)^2} 
\Bigg[ \, L_9 + L_{10}   + \left( C_{78} - \frac{5}{2}C_{88} - C_{89} 
+ 3C_{90} \right) p^2 \nl
&&  \hspace*{-2cm} \left. + \left( C_{78} - 2C_{87} + 
\frac{1}{2}C_{88}\right) q^2 
 + \left(C_{78} + 4C_{82} - \frac{1}{2}C_{88}\right)
(p+q)^2 \right] \,+ \cO(p^8) , \nn \\ 
{\cal G}^{\mrm{CHPT}}(p^2,q^2,(p+q)^2) & = & \frac{4 
\langle{\overline\psi}\psi\rangle_0}{F^2 q^2 (p+q)^2} 
\left[ L_9 + 2( -C_{88} +  C_{90}) p^2  \right. \nn\\ 
&&  \left. + (2C_{78} - C_{89} + C_{90}) q^2  - 2  C_{90} (p+q)^2
\right] \,+ \cO(p^8) \,.
\end{eqnarray}

\paragraph{3.}  

We next recall the properties of the $\VAP$ correlator at short
distances. Following Refs.~\cite{Moussallam:1997xx,Knecht:2001xc}, 
the analysis will be restricted to the leading term in the OPE.
Being an order parameter of chiral symmetry
breaking, the behaviour of the $\VAP$ Green function at short
distances is smoother than expected from naive power counting
arguments. The leading contributions at large momenta are proportional
to the quark condensate, which has the same anomalous dimension
as the pseudoscalar current. Hence the corresponding Wilson coefficient
in the OPE of the $\VAP$ correlator does not have an anomalous
dimension and, therefore, QCD corrections should modify the following 
results only mildly.
\par
As shown by Knecht and Nyffeler \cite{Knecht:2001xc}, two
short-distance limits are of interest here. In the first case, the two
momenta $p$ and $q$ in the correlator (\ref{VAPdef}) become 
simultaneously large. In position space this amounts to the
situation where the space-time arguments of the three operators
tend towards the same point at the same rate ($x \sim y \sim 0$). 
Restricting the discussion as always to the leading term in the OPE, 
one obtains \cite{Moussallam:1997xx,Knecht:2001xc}
\begin{eqnarray}
\label{ShortDistanceGeneral}
\lim_{\lambda\to\infty}(\Pi_{V\!AP})_{\mu\nu}^{abc}(\lambda p,\lambda q)
& = & \frac{\langle{\overline\psi}\psi\rangle_0}{\lambda^2}\,f^{abc}\,
\frac{1}{p^2q^2(p+q)^2}\,\bigg\{\,
p^2(p+2q)_{\mu}q_{\nu} \nn\\* 
&& -g_{\mu\nu}p^2q^2 +\frac{1}{2}(p^2-q^2-(p+q)^2)P_{\mu\nu}-Q_{\mu\nu}\,
\bigg\} +\,{\cal O}\left(\frac{1}{\lambda^4}\right)\,
\end{eqnarray}
and therefore
\begin{eqnarray}
\label{ShortDistanceFG}
\lim_{\lambda \to \infty} {\cal F}((\lambda p)^2, (\lambda q)^2, (\lambda p
+\lambda q)^2) &=& \frac{\langle{\overline\psi}\psi\rangle_0}
{2\lambda^4} \, 
\frac{p^2 - q^2 - (p+q)^2}{p^2 q^2 (p+q)^2} + 
\order\left(\frac{1}{\lambda^6}\right) \,, \nn\\
\lim_{\lambda \to \infty} {\cal G}((\lambda p)^2, (\lambda q)^2, (\lambda p
+ \lambda q)^2) &=& - \frac{\langle{\overline\psi}\psi\rangle_0}
{\lambda^6} \,\frac{1}{p^2 q^2 (p+q)^2} + 
 \order\left(\frac{1}{\lambda^8}\right) \,.
\end{eqnarray}
\par
The second situation of interest corresponds to the case where the
relative distance between only two of the three operators involved
becomes small. We refer to Ref.~\cite{Knecht:2001xc} for a complete
discussion of the various cases where different two-point functions 
arise. It turns out that many of the resulting conditions are not
independent when taken together with the constraint 
(\ref{ShortDistanceGeneral}). Therefore, we only reproduce the
following short-distance condition from Ref.~\cite{Knecht:2001xc},
which, together with (\ref{ShortDistanceGeneral}), leads to a complete
set of leading-order high-energy constraints: 
\begin{eqnarray}
\label{ShortDistance1}
\lim_{\lambda \to \infty}
(\Pi_{V\!AP})_{\mu\nu}^{abc}(\lambda p,q-\lambda p) &=& 
-\,\frac{1}{\lambda}\,f^{abc}\,\langle{\overline\psi}\psi\rangle_0
\,\frac{p_{\mu}q_{\nu}+p_{\nu}q_{\mu}-(p\cdot q)g_{\mu\nu}}{p^2 q^2}
+ \, {\cal O}\left(\frac{1}{\lambda^2}\right)\,.
\end{eqnarray}
In terms of the invariant functions ${\cal F}$ and ${\cal G}$, this 
asymptotic behaviour implies 
\begin{eqnarray}
\label{FGShortDistance1}
\lim_{\lambda \to \infty}
{\cal F}((\lambda p)^2,(q-\lambda p)^2,q^2) &=&
\frac{\langle{\overline\psi}\psi\rangle_0}{\lambda^2 p^2}\, 
\left[{\cal F}^{(0)}(q^2)\,+\, \frac{1}{\lambda}\,\frac{p\cdot q}{p^2}
\,{\cal F}^{(1)}(q^2)
\,+\,\order\left({1\over \lambda^2}\right)\right] \,, \nn\\
\lim_{\lambda \to \infty}
{\cal G}((\lambda p)^2,(q-\lambda p)^2,q^2) &=&
\frac{\langle{\overline\psi}\psi\rangle_0}{(\lambda^2 p^2)^2}\, 
\left[{\cal G}^{(0)}(q^2)\,+\, \frac{1}{\lambda}\,\frac{p\cdot q}{p^2}
\,{\cal G}^{(1)}(q^2)
\,+\,\order\left({1\over \lambda^2}\right)\right] \,, 
\end{eqnarray}
with
\begin{eqnarray}
\label{FGShortDistance2}
{\cal F}^{(0)}(q^2)-{\cal G}^{(0)}(q^2) = 
\displaystyle\frac{1}{q^2} \, , \quad & \quad
{\cal F}^{(1)}(q^2)-{\cal G}^{(1)}(q^2)+{\cal G}^{(0)}(q^2)= 
\displaystyle\frac{2}{q^2}\,.
\end{eqnarray}
\par
The $\VAP$ correlator is also related to the $\Gamma_{\rm VA}$ and
$\Gamma_{\rm VP}$ vertex functions \cite{Knecht:2001xc,Moussallam:1997xx}.
The short-distance behaviour of these
vertex functions gives additional constraints on the parameters of 
${\cal F}$ and ${\cal G}$. However, using the expansion of 
$\Gamma_{\rm VA}$ and $\Gamma_{\rm VP}$ consistently up to 
${\cal O}(1/\lambda)$,
those constraints are equivalent to the limit where one momentum of $\VAP$ 
becomes large and therefore they do not provide new information at leading
order.
\par
In order to solve the short-distance conditions
(\ref{ShortDistanceGeneral}) and (\ref{ShortDistance1}), we propose
the following ansatz inspired by large $N_C$ that is a generalization
of the one used in Refs.~\cite{Moussallam:1997xx,Knecht:2001xc}:
\begin{eqnarray}
 \label{FGansatz}
{\cal F}(p^2,q^2,(p+q)^2) &=&
\frac{\langle{\overline\psi}\psi\rangle_0}{(p^2-M_V^2)(q^2-M_A^2)}\nn\\*
&&\times \left[a_0+\frac{b_1+b_2 p^2 + b_3 q^2}{(p+q)^2}
+\frac{c_1 + c_2 p^2 + c_3 q^2}{(p+q)^2-M_P^2}\right] ~,\nn\\*
{\cal G}(p^2,q^2,(p+q)^2) &=&
\frac{\langle{\overline\psi}\psi\rangle_0}{(p^2-M_V^2)q^2}
\Bigg[\frac{d_1 + d_2 q^2}{(p+q)^2(q^2-M_A^2)}
+ \frac{f}{(p+q)^2-M_P^2}\Bigg] ~.
\end{eqnarray}
This ansatz differs from the one in
Refs.~\cite{Moussallam:1997xx,Knecht:2001xc} by
the inclusion of a nonet of pseudoscalar resonances with mass $M_P$
(remember that we always work in the chiral limit). The ansatz of 
Refs.~\cite{Moussallam:1997xx,Knecht:2001xc} is recovered in the 
limit $M_P \to \infty$, i.e. by dropping the parameters 
$c_1,c_2,c_3,f$ that specify the
contributions from pseudoscalar resonance exchange.
 While the ansatz of Refs.~\cite{Moussallam:1997xx,Knecht:2001xc} 
was designed to match the leading short-distance constraints with the
minimal resonance content, we include all lowest-lying resonance
multiplets that can contribute to the  LECs of ${\cal O} (p^6)$.
Our approach appears more natural when
attempting to construct an explicit Lagrangian realization for the 
resonance interactions (see discussion below).
\par
The parameters in (\ref{FGansatz}) fall into two classes:
\begin{itemize} 
\item[1)] The dimensionless parameters $a_0,b_2,b_3,c_2,c_3,d_2,f$ are
constrained by the OPE conditions (\ref{ShortDistanceGeneral}) 
and (\ref{ShortDistance1}).
\item[2)] The parameters $b_1,c_1,d_1$ with squared mass dimension are not
  affected by the (leading-order) OPE conditions. As discussed in the 
  next paragraph,
  they can be constrained  by asymptotic
  conditions on various form factors when one or two pions are put
  on-shell \cite{Moussallam:1997xx,Knecht:2001xc}.
\end{itemize} 

The short-distance conditions (\ref{ShortDistanceGeneral}) and 
(\ref{ShortDistance1}) yield the following set of six linear equations
for the seven parameters in class 1:
\begin{equation} 
 \label{OPEeq}
\begin{tabular}{lll} 
\quad $a_0 = -\displaystyle\frac{1}{2}$ , \qquad & \quad $b_2+c_2 =
\displaystyle\frac{1}{2}$ , \qquad & \quad 
$b_3+c_3 = -\displaystyle\frac{1}{2}$ , \nn \\[.2cm] 
\quad $d_2+f = -1$ , \qquad & \quad $b_2+b_3-d_2 = 1$ ,
\qquad & \quad $2 b_2-d_2 = 2$ .
\end{tabular} 
\end{equation}  
We use these equations to express six of the parameters in
terms of $b_3$:
\begin{equation}
\label{OPEsol} 
\begin{tabular}{lll} 
$a_0 = -\displaystyle\frac{1}{2}$ ,\hspace*{1cm}  & $b_2 = 1 + b_3$ , 
\hspace*{1cm} & $d_2 = 2 b_3$ , \\[.2cm]  
$c_2 = -\displaystyle\frac{1}{2} - b_3$ , \hspace*{1cm}  & 
$c_3 = c_2$ ,
\hspace*{1cm} & $f = 2 c_2$ .
\end{tabular} 
\end{equation} 
Setting the pseudoscalar exchange parameters $c_2,c_3$ and $f$ to zero
(or, equivalently, letting $M_P \to \infty$ in the ansatz
(\ref{FGansatz})), we 
recover the solution of Refs.~\cite{Moussallam:1997xx,Knecht:2001xc}:
\begin{equation} 
\label{KNsol}
\begin{tabular}{llll} 
$a_0 = -\displaystyle\frac{1}{2}$ ,\hspace*{1cm}  & 
$b_2 = \displaystyle\frac{1}{2}$ , \hspace*{1cm} &  
$b_3 = -\displaystyle\frac{1}{2}$ , \hspace*{1cm} &
$d_2 = - 1$ .
\end{tabular} 
\end{equation}

\paragraph{4.}  
Additional information on the parameters in  Eq.~(\ref{FGansatz})
can be obtained by putting one or two momenta in the $\VAP$ Green
function (\ref{VAPdef}) on the pion mass shell
\cite{Moussallam:1997xx,Knecht:2001xc}. The form factors appearing in 
the resulting vertex functions are not directly constrained by QCD but 
there are strong theoretical arguments \cite{Lepage:1979zb,Lepage:1980fj,
Brodsky:1981rp} for the form factors in question to fall off at least
like $1/q^2$ for large momentum transfers.

The two dimensional parameters $b_1,d_1$ were determined in
this way by Moussallam \cite{Moussallam:1997xx}, making also use of
the two Weinberg sum rules \cite{Weinberg:1967kj}. The inclusion of 
pseudoscalar resonances does not affect those results:
\begin{eqnarray}
\label{b1d1} 
b_1 = M_A^2 - M_V^2 ~, \qquad & \qquad d_1 = 2 M_A^2 ~.
\end{eqnarray}

Although we cannot determine $c_1$ from consideration of a pionic form
factor we can fix the remaining dimensionless parameter $b_3$
in this way. For this purpose, we consider the axial form factor
$G_A(t)$ governing the matrix element 
$\langle \gamma | A_{\mu} | \pi \rangle$ \cite{Gasser:1984yg}.
Extracting $G_A(t)$ from the
Green function (\ref{VAPdef}) by setting $p^2=0$ and $(p+q)^2=0$
(massless pion), one finds in terms of the parameters defined in 
(\ref{FGansatz}) 
\begin{equation}
\label{ganew} 
G_A(t) = \Frac{F^2}{M_V^2} \; \Frac{b_1+b_3 t}{M_A^2-t} ~.
\end{equation} 
Demanding that the form factor $G_A(t)$ vanishes for large $t$ 
\cite{Lepage:1979zb,Lepage:1980fj,Brodsky:1981rp,Ecker:1989yg}, we
obtain
\begin{equation}
\label{b3} 
b_3=0~.
\end{equation}
Therefore, the solution (\ref{KNsol})
\cite{Moussallam:1997xx,Knecht:2001xc} 
is not compatible with the asymptotic
vanishing of $G_A(t)$. The value $b_3= -1/2$ in (\ref{KNsol})
is also at the origin
of the very small partial width obtained 
\cite{Moussallam:1997xx,Knecht:2001xc} for the decay $a_1 \to
\pi \gamma$. The decay matrix element is governed 
by the combination $b_1 + b_3 M_A^2$. With the solution (\ref{KNsol}),
this matrix element is proportional to 
$(M_A^2 - 2 M_V^2)/2$ and therefore suppressed 
compared to our solution with $b_3=0$ where the same matrix element is 
given by $b_1= M_A^2 - M_V^2$. The numerical value of the decay width
$\Gamma(a_1 \to \pi \gamma)$ will be discussed later.
\par
 The $\VAP$ Green function also contributes to the decay
$\tau \rightarrow 3 \pi \nu_{\tau}$. One can study the
axial-vector form factor contributing to this process and require that
it vanishes like $1/q^2$ for large momentum transfer. This procedure
provides the conditions
\begin{equation}
b_2 = 1 \; , \; \; \; \; \;  \; \; \; b_3=0 \; , \;\; \; \; \; 
 \; \; \; d_2 = 0 
\end{equation}
that are consistent with the results in Eqs.~(\ref{OPEsol}) and
(\ref{b3}). 
\par
 The discussion above is related to a general point raised 
in Ref.~\cite{Bijnens:2003rc}. 
There it was claimed that for a given three-point function 
that is an order parameter of chiral symmetry breaking,
any large-$N_C$ inspired ansatz with a {\em finite number} of resonance
multiplets will fail to reproduce simultaneously (i) the leading OPE
constraints and (ii) the $1/q^2$ asymptotic behaviour of  {\em all}
hadronic form factors (appearing as residues of two-particle poles in
the Green function).  
Our explicit construction shows that with a reasonable number of
resonance multiplets one can still fulfill both the leading OPE 
constraints and the correct asymptotic behaviour of form factors 
involving Goldstone modes and on--shell photons 
($F_V^{\pi}(t)$ and $G_{A}(t)$ in our case).

\paragraph{5.}  

We now turn to an explicit realization of our solutions
(\ref{OPEsol}), (\ref{b1d1}) and (\ref{b3}) in terms of a chiral 
resonance Lagrangian. Such a Lagrangian was introduced in 
Refs.~\cite{Ecker:1989te,Ecker:1989yg} to investigate the LECs of
$\cO(p^4)$. That Lagrangian has to be extended when going up to
$\cO(p^6)$ where bilinear resonance couplings also contribute.

In the notation of Refs.~\cite{Ecker:1989te,Ecker:1989yg}, the
kinetic terms of the Lagrangian restricted to vector, axial-vector and
pseudoscalar resonance fields ($V(1^{--})$, $A(1^{++})$ and
$P(0^{-+})$) are given by
\begin{eqnarray}
{\cal L}_{ \rm kin}^R &=& -\frac{1}{2} \langle
\nabla^\lambda R_{\lambda\mu} \nabla_\nu R^{\nu\mu} - \frac{M_R^2}{2}
R_{\mu\nu} R^{\mu\nu} \rangle \; \; \; , \; \; \;  \; \; R \, = \, V,A 
\; , \nn \\
{\cal L}_{ \rm kin}^P &=& \frac{1}{2} \langle \nabla^{\mu} P \nabla_{\mu} P
- M_P^2 P^2 \rangle \; .
\end{eqnarray}
$V_{\mu\nu}$ and $A_{\mu\nu}$ are antisymmetric tensor fields
describing nonets of spin-1 mesons and $P$ is a pseudoscalar
(nonet) field. The brackets $\langle \dots \rangle$ denote a 
three-dimensional trace in flavour space. In the large-$N_C$ limit
where multiple trace terms are suppressed, the interaction terms
linear in the resonance fields  and with the minimal number of
derivatives and mass insertions are given by \cite{Ecker:1989te}
\begin{eqnarray}
\label{lag1}
{\cal L}_2^{V,A,P} & = &  \frac{F_V}{2\sqrt{2}} \langle V_{\mu\nu}
f_+^{\mu\nu}\rangle + i\,\frac{G_V}{\sqrt{2}} \langle V_{\mu\nu} u^\mu
u^\nu\rangle +  \frac{F_A}{2\sqrt{2}} \langle A_{\mu\nu}
f_-^{\mu\nu}\rangle + i \, d_{m} \, \langle P \chi_{-} \rangle ~.
\end{eqnarray}
The chiral fields $u^\mu,f_+^{\mu\nu},f_-^{\mu\nu},\chi_{-}$ are
defined as usual \cite{Ecker:1989te,Ecker:1989yg} in terms of
Goldstone fields and external fields. The coupling
constants $F_V$, $G_V$, $F_A$ and $d_m$ are real.

To account for LECs of $\cO(p^6)$, we must also include couplings of 
Goldstone bosons with two resonance fields. As for the linear
couplings (\ref{lag1}), we only include terms with the minimal number
of derivatives and mass insertions. 
Although there is a priori no guarantee that the various
asymptotic constraints discussed previously can be satisfied with such 
a minimal Lagrangian this approach proved to be successful at 
$\cO(p^4)$ \cite{Ecker:1989te,Ecker:1989yg}.

The vector--axial-vector bilinear terms were already
introduced in Ref.~\cite{Dumm:2003ku}:\\
\begin{equation} 
\label{OVA}
\begin{tabular}{lll} 
${\cal L}_2^{VA}  =  \sum_{i=1}^{5} \, \lambda_i^{VA} \, 
{\cal O}_i^{VA}$  & \hspace*{1cm} &  \\[.2cm] 
${\cal O}_1^{VA} =  \langle \,  [ \, V^{\mu\nu} \, , \, 
A_{\mu\nu} \, ] \,  \chi_- \, \rangle$ , & &
${\cal O}_2^{VA}  =  i\,\langle \, [ \, V^{\mu\nu} \, , \, 
A_{\nu\alpha} \, ] \, h_\mu^{\;\alpha} \, \rangle$ , \\[.2cm] 
${\cal O}_3^{VA} =   i \,\langle \, [ \, \nabla^\mu V_{\mu\nu} \, , \, 
A^{\nu\alpha}\, ] \, u_\alpha \, \rangle$ , & &
${\cal O}_4^{VA}  =  i\,\langle \, [ \, \nabla_\alpha V_{\mu\nu} \, , \, 
A^{\alpha\nu} \, ] \,  u^\mu \, \rangle$ , \\[.2cm] 
${\cal O}_5^{VA} =   i \,\langle \, [ \, \nabla_\alpha 
V_{\mu\nu} \, , \, 
A^{\mu\nu} \, ] \, u^\alpha \, \rangle$ .& & \\[.1cm] 
\end{tabular} 
\end{equation}
The pseudoscalar--vector bilinear couplings are given by \\[.1cm] 
\begin{equation} 
\label{OPV}
\begin{tabular}{lll} 
${\cal L}_2^{PV} =  \sum_{i=1}^{2} \, \lambda_i^{PV} \, 
{\cal O}_i^{PV}$  & \hspace*{1cm} & \\[.2cm] 
${\cal O}_1^{PV} =  i \, \langle \,  [ \, \nabla^\mu P \, , \, 
V_{\mu\nu} \, ] \,  u^\nu \, \rangle$ , & &
${\cal O}_2^{PV}   =  i \, \langle \,  [ \, P \, , \, 
V_{\mu\nu} \, ] \,  f^{\mu\nu}_- \, \rangle$ .\\[.1cm] 
\end{tabular} 
\end{equation} 
Finally, only one bilinear term with pseudoscalar and axial-vector 
fields contributes to our Green function:
\begin{eqnarray}
\label{OPA}
{\cal L}_2^{PA} & = &  \lambda_1^{PA} \, 
{\cal O}_1^{PA}  \nl
{\cal O}_1^{PA} &  = & i \, \langle \,  [ \, P \, , \, 
A_{\mu\nu} \, ] \,  f^{\mu\nu}_+ \, \rangle \,.
\end{eqnarray}
Adding the lowest-order chiral Lagrangian \cite{Gasser:1985gg}, 
we obtain the following chiral resonance Lagrangian to be used for the 
calculation of the $\VAP$ Green function:
\begin{equation}
\label{TotalLagrangian}
{\cal L}_{\mrm{CHRL}} \, = \, \frac{F^2}{4}\langle u_{\mu}
u^{\mu} + \chi _+ \rangle  + 
{\cal L}_{ \mrm{kin}}^{V,A,P}  +  {\cal L}_2^{V,A,P}  +  
{\cal L}_2^{VA}  +  {\cal L}_2^{PV}  + 
{\cal L}_2^{PA} \; .
\end{equation}

\paragraph{6.}  
\begin{figure}
\begin{center}
\includegraphics[height=100pt,width=100pt,angle=0]{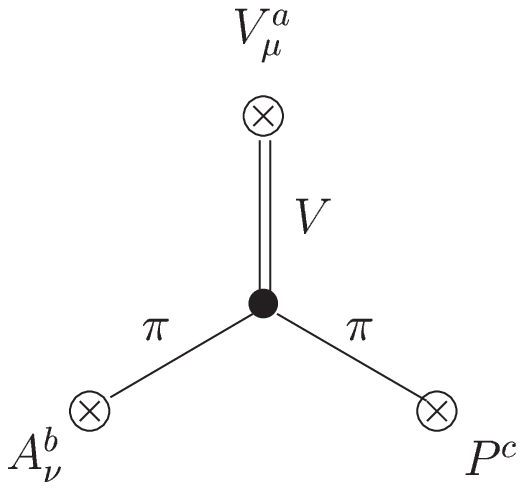}
\hspace*{0.5cm}
\includegraphics[height=100pt,width=100pt,angle=0]{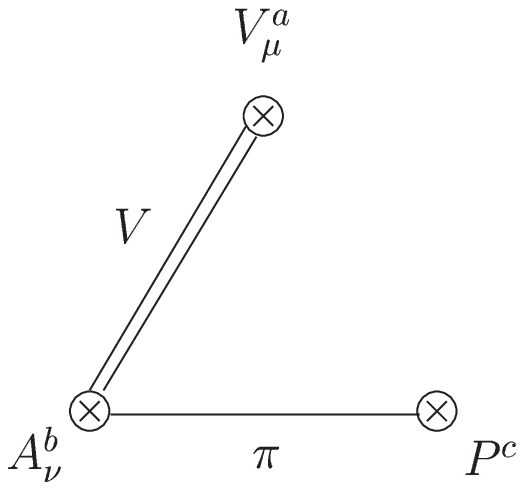}
\hspace*{0.5cm}
\includegraphics[height=100pt,width=100pt,angle=0]{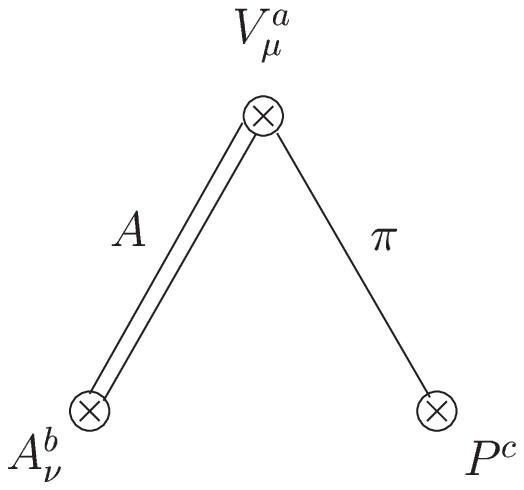}
\caption{\label{fig:OneResonance}
Single-resonance exchange: $\pi$ denotes a Goldstone boson, $V$ and 
$A$ stand for vector and axial-vector resonances, respectively.}
\end{center}
\end{figure}
Lowest-order Goldstone boson exchange provides
the first two terms in the $\VAP$ correlator (\ref{GeneralSolution}) 
that drive the chiral Ward identities (\ref{WardIdentities}).
Next we compute the diagrams shown in 
Fig.~\ref{fig:OneResonance} where a single resonance is exchanged. 
There is no contribution from pseudoscalar resonance exchange in this
case \cite{Ecker:1989te}. 
The diagrams in Fig.~\ref{fig:OneResonance} give rise to 
\begin{eqnarray}
 \label{L1FGResults}
{\cal F}^{V,A}(p^2,q^2,(p+q)^2) &=& 
\frac{\langle{\overline\psi}\psi\rangle_0}{(p+q)^2 (p^2-M_V^2)}
\left(\frac{F_V^2-2 F_V G_V}{F^2}
-  \frac{p^2-M_V^2}{q^2-M_A^2}\frac{F_A^2}{F^2}\right)~, \nn\\
{\cal G}^{V,A}(p^2,q^2,(p+q)^2) &=& 
\frac{\langle{\overline\psi}\psi\rangle_0}{q^2 (p+q)^2 (p^2-M_V^2)} 
\frac{- 2 F_V G_V}{F^2} \,.
\end{eqnarray}

\begin{figure}
\begin{center}
\includegraphics[height=100pt,width=100pt,angle=0]{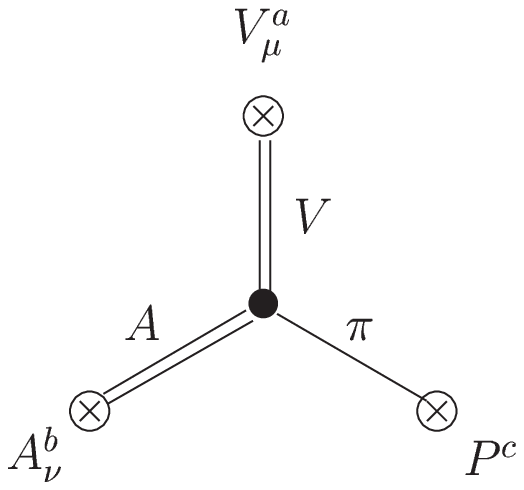}
\hspace*{0.5cm}
\includegraphics[height=100pt,width=100pt,angle=0]{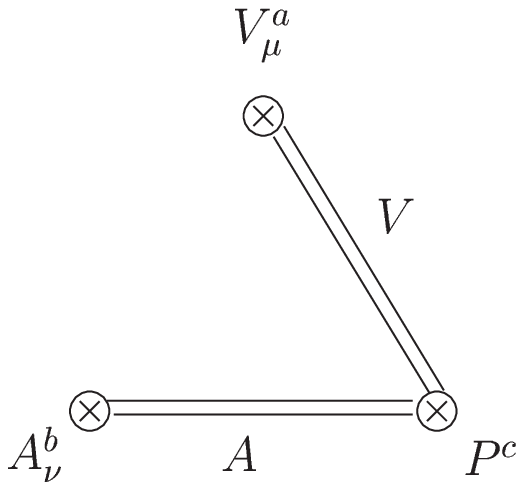}
\hspace*{0.5cm}
\includegraphics[height=100pt,width=100pt,angle=0]{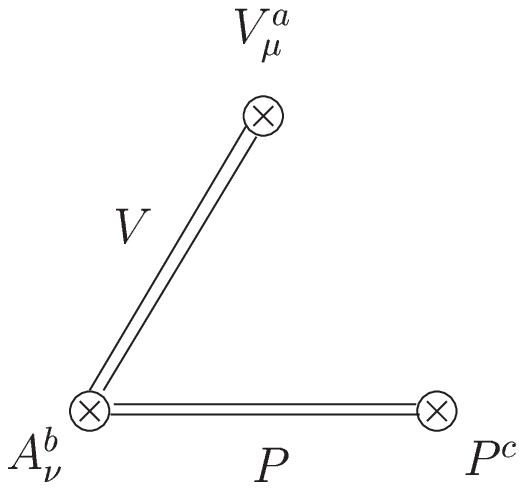}\\[4mm]
\includegraphics[height=100pt,width=100pt,angle=0]{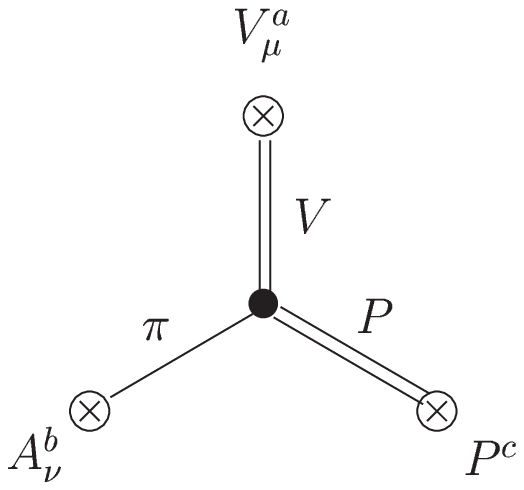}
\hspace*{0.5cm}
\includegraphics[height=100pt,width=100pt,angle=0]{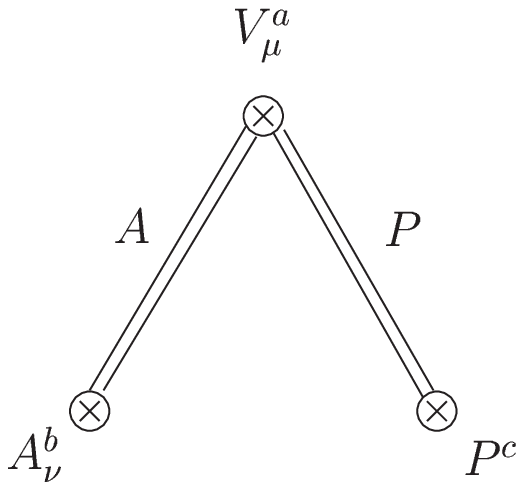}
\caption{\label{fig:TwoResonances}
Double-resonance exchange contribution to the $\VAP$ Green function; 
$P$ denotes a pseudoscalar resonance.}
\end{center}
\end{figure}

The double-resonance contributions to the $\VAP$ Green function are
described by the diagrams in Fig.~\ref{fig:TwoResonances}.
Summing up single- and double-resonance exchange contributions, the
final result can be given in terms of the parameters defined in the
general ansatz (\ref{FGansatz}):
\begin{equation} 
 \label{FGParameters}
\begin{tabular}{ll} 
$a_0 =  -2 \sqrt{2}\displaystyle\frac{F_V F_A}{F^2}\lambda_0$ ,  &
$b_1 = M_V^2\displaystyle\frac{F_A^2}{F^2}-M_A^2 
\displaystyle\frac{F_V^2-2F_V G_V}{F^2}$ , \nn\\
$b_2 =  -\displaystyle\frac{F_A^2}{F^2}+
2\sqrt{2}\displaystyle\frac{F_V F_A}{F^2}\lambda'$ , &
$b_3 =  \displaystyle\frac{F_V^2-2F_V G_V}{F^2} +
2\sqrt{2}\displaystyle\frac{F_V F_A}{F^2}\lambda''$ , \nn\\
$c_1 = -M_V^2 c_2 - M_A^2 c_3$ , &
$c_2 = - 8\sqrt{2}\,\displaystyle\frac{F_A d_m}{F^2}\, \lambda_1^{PA}$ , \nn\\
$c_3 = 8\sqrt{2}\,\displaystyle\frac{F_V d_m}{F^2}\, 
\left(\displaystyle\frac{\lambda_1^{PV}}{2}+\lambda_2^{PV}\right)$ , &
$d_1 = \displaystyle\frac{2 F_V G_V}{F^2} M_A^2$ , \nl
$d_2 = - \displaystyle\frac{2 F_V G_V}{F^2} + 
2\sqrt{2}\, \displaystyle\frac{F_V F_A}{F^2}(\lambda'+\lambda'')$ ,
\qquad  &
$f = 4\sqrt{2}\,\displaystyle\frac{F_V d_m}{F^2}\, \lambda_1^{PV}$~,
\end{tabular} 
\end{equation} 
where we have used the definitions \cite{Dumm:2003ku}
\begin{eqnarray}
 \label{LambdaDef}
\sqrt{2} \lambda_0 &=& -4\lambda_1^{VA}-\lambda_2^{VA}
-\frac{\lambda_4^{VA}}{2}-\lambda_5^{VA}~, \nn\\
\sqrt{2} \lambda' &=& \lambda_2^{VA}-\lambda_3^{VA}
+\frac{\lambda_4^{VA}}{2}+\lambda_5^{VA}~, \nn\\
\sqrt{2} \lambda'' &=& \lambda_2^{VA}
-\frac{\lambda_4^{VA}}{2}-\lambda_5^{VA}~.
\end{eqnarray}

As pointed out in  Ref.~\cite{Ecker:1989yg}, the short-distance
structure of QCD can be used to constrain the couplings of the 
chiral resonance Lagrangian by matching the asymptotic 
behaviour of two-point functions, form factors and scattering 
amplitudes with the results from resonance exchange (at leading 
order in the $1/N_C$ expansion and assuming a single nonet of $V$ 
and $A$ resonances each). In this way one finds
the relations \cite{Ecker:1989yg,Weinberg:1967kj} 
\begin{eqnarray} 
\label{FVGVFA}
F_V G_V = F^2, \qquad & F_V^2 - F_A^2 = F^2 , \qquad &
F_V^2 M_V^2 = F_A^2 M_A^2~,
\end{eqnarray} 
allowing to express the couplings $F_V$, $G_V$ and $F_A$ in terms of
$F$, $M_V$ and $M_A$. From a similar joint analysis of the scalar form factor 
\cite{Jamin:2000wn,Jamin:2001zq} and the SS-PP sum rules 
\cite{Golterman:1999au} one gets, assuming again only one nonet of 
$S$ and $P$ resonances each \cite{Pich:2002xy},
\begin{equation}
\label{dm}
d_m = \Frac{F}{2 \sqrt{2}} \; \; .
\end{equation}

The ten relations in (\ref{FGParameters}) can now be compared with the
previous results (\ref{OPEsol}), (\ref{b1d1}) and (\ref{b3}). From the
equations for $a_0,b_2,b_3$ we extract the combinations of 
coupling constants $\lambda_0,\lambda',\lambda''$ that satisfy the 
relation $4 \lambda_0 = \lambda' + \lambda''$.
The equation for $d_2$ is then automatically satisfied.
The relations for $c_2,c_3,f$ fix the coupling
constants $\lambda_1^{PV},\lambda_2^{PV},\lambda_1^{PA}$. The
equations for $b_1,d_1$ are consistent with (\ref{b1d1}) and there is
a new relation for the dimensional parameter $c_1$:
\begin{equation}
\label{c1} 
c_1 = \displaystyle\frac{1}{2} (M_V^2 + M_A^2)~. 
\end{equation}
 
The predictions of the chiral resonance Lagrangian 
(\ref{TotalLagrangian}) are fully consistent with the OPE and form
factor constraints (\ref{OPEsol}), (\ref{b1d1}) and (\ref{b3}). In
other words, at leading order in $1/N_C$ and considering for each
current in the Green function only one multiplet of resonances with
the same quantum numbers, the chiral resonance
Lagrangian provides a $\VAP$ Green function with the correct
asymptotic behaviour dictated by QCD.

Using (\ref{FVGVFA}) and (\ref{dm}), we get the following final
results for the coupling constants of the chiral Lagrangians 
(\ref{OVA}), (\ref{OPV}) and (\ref{OPA}), depending only on the masses
$M_V$ and $M_A$:
\begin{equation}
\label{lambda}
\begin{tabular}{lclcl}
$\lambda' = \displaystyle\frac{M_A}{2 \sqrt{2} M_V}$ , & \hspace*{1cm} &    
$\lambda'' = \displaystyle\frac{M_A^2 - 2 M_V^2}{2 \sqrt{2} M_V M_A}$ ,
  & \hspace*{1cm} & $4 \lambda_0 = \lambda' + \lambda''$ , \nn \\[.4cm] 
$\lambda_1^{PV}= - 4 \lambda_2^{PV}$ , & \hspace*{1cm} &
$\lambda_2^{PV} = \displaystyle\frac{\sqrt{M_A^2 - M_V^2}}{8 M_A}$ , &
\hspace*{1cm} & 
$\lambda_1^{PA} = \displaystyle\frac{\sqrt{M_A^2 - M_V^2}}{8 M_V}$~.
\end{tabular}
\end{equation}

We now come back to the axial form factor $G_A(t)$ in the matrix 
element $\langle \gamma | A_{\mu} | \pi \rangle$. With
single-resonance exchange only, this form factor is given by 
\cite{Ecker:1989yg}
\begin{equation}
\label{gaold}
G_A(t) = \Frac{2F_V G_V - F_V^2}{M_V^2} + \Frac{F_A^2}{M_A^2-t} \; \;.
\end{equation}
Requiring $G_A(t)$ to vanish for $t \to \infty$ implies the relation
$F_V = 2 G_V$, one version of the so-called KSFR relation 
\cite{Kawarabayashi:1966kd,Riazuddin:1966sw}. The inclusion of 
bilinear resonance couplings modifies the form factor as given
in Eq.~(\ref{ganew}) with $b_1=M_A^2 - M_V^2, ~b_3=0$, and it induces
a correction to the KSFR relation:
\begin{eqnarray} 
\displaystyle\frac{2 F_V G_V - F_V^2}{2 F^2} = 1 -
\displaystyle\frac{F_V^2}{2 F^2} = \displaystyle\frac{M_A^2 - 2 M_V^2}
{2(M_A^2 - M_V^2)}  \; \; .
\end{eqnarray}
With $M_A=1.23$ GeV, $M_V=0.771$ GeV \cite{Hagiwara:2002fs}, the 
right-hand side takes the value $\simeq 0.18$. The
partial decay width  $\Gamma(a_1 \to \pi \gamma)$ is now 
\begin{equation}
\label{width}
 \Gamma(a_1 \to \pi \gamma) = \displaystyle\frac{\alpha M_A}{24}
  \left( \displaystyle\frac{M_A^2}{M_V^2} - 1 \right)^3
 \left(1 - \displaystyle\frac{M_\pi^2}{M_A^2} \right)^3~. 
\end{equation}
With the same values for $M_V,M_A$ and with the physical pion mass,
we obtain $\Gamma(a_1 \to \pi \gamma) = 1.33$ MeV, in reasonable
agreement with the experimental value $640 \pm 246$ keV 
\cite{Zielinski:1984au}. 
It should be noted that the width is very sensitive to $M_A$. 
A value of $M_A=\sqrt{2}M_V$, as required by the KSFR relation, 
would give $\Gamma(a_1 \to \pi \gamma) = 316$ keV
and for $M_A=1.2$ GeV, a value extracted from  $\tau\to 3 \pi\nu_\tau$
data \cite{Dumm:2003ku}, one obtains $\Gamma(a_1 \to \pi \gamma) = 1.01$ MeV.
In comparison, the decay
width in the scenario of Refs.~\cite{Moussallam:1997xx,Knecht:2001xc}
is strongly suppressed with respect to (\ref{width}) by a factor
\begin{equation} 
\displaystyle\frac{(M_A^2 - 2 M_V^2)^2}{4(M_A^2 - M_V^2)^2} \simeq
0.03 ~.
\end{equation} 
Although the corresponding width is more than an order of magnitude
smaller than the listed value \cite{Hagiwara:2002fs} the experimental 
situation (discussed in Ref.~\cite{Moussallam:1997xx}) remains to be 
settled. However, we are confident that future experiments will be
able to decide between two predictions that differ by a factor 30.
\par
As commented above, the relations (\ref{lambda}) also have a bearing on 
the decays
$\tau \to  3 \pi \nu_\tau$. As shown in Ref.~\cite{Dumm:2003ku}, the
requirement that the $J=1$ axial spectral function vanishes for large
momentum transfer implies certain values for $\lambda'$,
$\lambda''$. Those values coincide\footnote{In Ref.~\cite{Dumm:2003ku} 
the KSFR relation $F_V=2 G_V$ was adopted implying $M_A^2 = 2 M_V^2$.} 
with the corresponding results in Eq.~(\ref{lambda}). The 
coupling $\lambda_0$ was extracted in that reference from a fit to the 
spectrum and branching ratio of the decay. However, the fitted value 
turns out to be too large and, as discussed in that reference, carries 
a big uncertainty due to the fact that in the $\tau \rightarrow 3 \pi 
\nu_{\tau}$ amplitude the  coupling $\lambda_0$ always appears
multiplied by a factor $M_{\pi}^2$.

\paragraph{7.}  

Having established the compatibility between the QCD short-distance
constraints and the chiral resonance Lagrangian, we can now use 
the results (\ref{OPEsol}), (\ref{b1d1}) and (\ref{b3}) (with the 
additional relation (\ref{c1}) for
$c_1$) to compare with the low-energy expansion 
(\ref{ChPTSolution}) of the $\VAP$ Green function. It turns out that
all LECs appearing in (\ref{ChPTSolution}) can be determined
separately in this way:
\begin{equation} 
 \label{LowEnergySolution}
\begin{tabular}{lll} 
$L_9$ = $\displaystyle\frac{F^2}{2 M_V^2}$  , & \hspace*{.1cm} &
$L_{10}$ =  $- \displaystyle\frac{F^2 (M_A^2 + M_V^2)}{4 M_V^2 M_A^2}$ 
 ,  \\[.3cm] 
$C_{78} = \displaystyle\frac{F^2(3 M_A^2 + 4 M_V^2)}{8 M_V^4 M_A^2} 
- \displaystyle\frac{F^2}{16 M_V^2 M_P^2}$  , & \hspace*{.1cm} &
$C_{82} = - \displaystyle\frac{F^2(4 M_A^2 + 5 M_V^2)}{32 M_V^4
  M_A^2} 
- \displaystyle\frac{F^2}{32 M_A^2 M_P^2}$  ,  \\[.3cm] 
$C_{87} = \displaystyle\frac{F^2(M_A^4 + M_V^4 + M_A^2 M_V^2)}
{8 M_V^4 M_A^4}$  ,  & \hspace*{.1cm} &
$C_{88} = - \displaystyle\frac{F^2}{4 M_V^4} + 
\displaystyle\frac{F^2}{8 M_V^2 M_P^2}$  ,  \\[.3cm] 
$C_{89} = \displaystyle\frac{F^2(3 M_A^2 + 2 M_V^2)}{4 M_V^4 M_A^2}$ 
 ,  & \hspace*{.1cm} & 
$C_{90} = \displaystyle\frac{F^2}{8 M_V^2 M_P^2}$  .
\end{tabular} 
\end{equation} 

The results for $L_9$ and $L_{10}$, the LECs of $\cO(p^4)$, coincide 
with those in Ref.~\cite{Ecker:1989yg}. The LECs of $\cO(p^6)$
differ from the ones in Ref.~\cite{Knecht:2001xc}, first of all by terms
involving the mass $M_P$ of the pseudoscalar resonance nonet. However,
even in the limit $M_P \to \infty$ a  small difference remains 
for the LECs
$C_{78}$ and $C_{82}$. The reason for this difference is
that the short-distance limit and the limit $M_P \to \infty$ do not
commute, as it is evident from the analysis of Eq.~(\ref{OPEeq}). 
Since $M_P \simeq 1.3 \, \mbox{GeV}$ is rather heavy, the pseudoscalar
contributions to the LECs are not large, ranging from $0.02 \times 10^{-4}$
to $0.09 \times 10^{-4}$ for $F^2 \, C_i$. All contributions from 
pseudoscalar resonances originate from the $f$ parameter in 
Eq.~(\ref{FGansatz}), except for $C_{82}$ which also gets a contribution
from the $c_1$ term.
\par
Our large-$N_C$ determination of these LECs cannot reproduce their
scale dependence (a next-to-leading-order effect in the $1/N_C$ counting). 
We have checked that in all cases the variation of the renormalized
LECs between $\mu = M_K$ and $\mu = 1$ GeV remains within a 
range of $30 \,\%$. Although there is no reason a priori that this
result will be valid for all LECs our findings quantify the
reliability of the estimates (\ref{LowEnergySolution}) for 
phenomenological applications.

\paragraph{8.}  

We have extended the ansatz of Moussallam, Knecht and Nyffeler
\cite{Moussallam:1997xx,Knecht:2001xc} for the $\VAP$ Green function
in the intermediate energy region by including the lowest-lying nonet
of pseudoscalar resonances. Since the model has more parameters
it trivially satisfies all short-distance constraints discussed 
in Refs.~\cite{Moussallam:1997xx,Knecht:2001xc}. \\[.2cm] 
The distinctive features of our solution are the following:
\begin{itemize} 
\item The axial form factor in the matrix element for the 
  decay $\pi \to e \nu_e \gamma$ vanishes for $t \to \infty$.
\item We obtain a partial decay width $\Gamma(a_1 \to \pi \gamma)$ 
  in reasonable agreement with the experimental value 
  \cite{Zielinski:1984au} but more than a factor 30 bigger
  than the prediction of Refs.~\cite{Moussallam:1997xx,Knecht:2001xc}.
\item The asymptotic vanishing of the axial spectral function relevant 
  for the decay $\tau \to  3 \pi \nu_\tau$ \cite{Dumm:2003ku} is
  compatible with the model. 
\item The LECs of $\cO(p^6)$ determined by the matching procedure
  differ in general from the ones derived in Ref.~\cite{Knecht:2001xc}
  even in the limit $M_P \to \infty$ for the mass of the pseudoscalar
  resonance nonet. 
\item The solution of the short-distance constraints is consistent
  with a minimal chiral resonance Lagrangian with vector, axial-vector
  and pseudoscalar resonances where the spin-1 mesons are described
  by antisymmetric tensor fields.
\end{itemize}

\paragraph{Acknowledgements}
\noindent 
We wish to thank Roland Kaiser and Pedro D. Ruiz-Femenia for useful 
discussions on the topic of this paper. We also thank Eduardo de Rafael,
Marc Knecht, Bachir Moussallam, Andreas Nyffeler and Joaquim Prades for
comments on the manuscript.
V.C. is supported by a Sherman Fairchild Fellowship from Caltech. 
M.E. thanks the European Union for financial support under contract 
no. HPMF-CT-2001-01128. This work has been supported in part by 
MCYT (Spain) under grant FPA2001-3031 and by ERDF funds from the
European Commission.

\vspace*{1cm} 

\begin{thebibliography}{99}

\bibitem{Moussallam:1997xx} B.~Moussallam, {\em Nucl. Phys.} {\bf B504}
(1997) 381.

\bibitem{Knecht:2001xc} M.~Knecht and A.~Nyffeler, {\em Eur. Phys. J.} 
{\bf C21} (2001) 659.

\bibitem{MHA} S.~Peris, M.~Perrottet and E.~de Rafael, {\em JHEP} {\bf 05}
(1998) 011; \\
S.~Peris, B.~Phily and E.~de Rafael, {\em Phys. Rev. Lett.} {\bf 86} (2001)
14; \\
E.~de Rafael, {\em Nucl. Phys. Proc. Suppl.} {\bf 119} (2003) 71. 

\bibitem{Ecker:1989te} G.~Ecker, J.~Gasser, A.~Pich and E.~de~Rafael, 
{\em Nucl. Phys.} {\bf B321} (1989) 311.

\bibitem{Ecker:1989yg} G.~Ecker {\em et al.}, {\em Phys. Lett.} {\bf B223}
(1989) 425.

\bibitem{Prades:1994ys} J.~Prades, {\em Z. Phys.} {\bf C63} (1994) 491.

\bibitem{Ruiz-Femenia:2003hm} P.D.~Ruiz-Femen\'{\i}a, A.~Pich and J.~Portol\'es,
{\em JHEP} {\bf 07} (2003) 003.

\bibitem{Fearing:1996ga} H.W.~Fearing and S.~Scherer, {\em Phys. Rev.} 
{\bf D53} (1996) 315.

\bibitem{Bijnens:1999sh} J.~Bijnens, G.~Colangelo and G.~Ecker, {\em JHEP}
{\bf 02} (1999) 020.

\bibitem{Bijnens:2001bb} J.~Bijnens, L.~Girlanda and P.~Talavera, 
{\em Eur. Phys. J.} {\bf C23} (2002) 539.

\bibitem{Ebertshauser:2001nj} T.~Ebertshauser, H.W.~Fearing and S.~Scherer,
{\em Phys. Rev.} {\bf D65} (2002) 054033.

\bibitem{Bijnens:2003rc} J.~Bijnens, E.~G\'amiz, E.~Lipartia and J.~Prades,
{\em JHEP} {\bf 04} (2003) 055.

\bibitem{Lepage:1979zb} G.P.~Lepage and S.J.~Brodsky, {\em Phys. Lett.}
{\bf B87} (1979) 359.

\bibitem{Lepage:1980fj} G.P.~Lepage and S.J.~Brodsky, {\em Phys. Rev.}
{\bf D22} (1980) 2157.

\bibitem{Brodsky:1981rp} S.J.~Brodsky and G.P.~Lepage, {\em Phys. Rev.}
{\bf D24} (1981) 1808.

\bibitem{Gasser:1984yg} J.~Gasser and H.~Leutwyler, {\em Ann. Phys.} {\bf 158}
(1984) 142.

\bibitem{Weinberg:1967kj} S.~Weinberg, {\em Phys. Rev. Lett.} {\bf 18} (1967)
507.

\bibitem{Dumm:2003ku} D.~G\'omez Dumm, A.~Pich and J.~Portol\'es, 
{\em Phys. Rev.} {\bf D69} (2004) 073002.

\bibitem{Gasser:1985gg} J.~Gasser and H.~Leutwyler, {\em Nucl. Phys.} 
{\bf B250} (1985) 465.

\bibitem{Jamin:2000wn} M.~Jamin, J.A.~Oller and A.~Pich, {\em Nucl. Phys.}
{\bf B587} (2000) 331.

\bibitem{Jamin:2001zq} M.~Jamin, J.A.~Oller and A.~Pich, {\em Nucl. Phys.}
{\bf B622} (2002) 279.

\bibitem{Golterman:1999au} M.F.L.~Golterman and S.~Peris, {\em Phys. Rev.}
{\bf D61} (2000) 034018.

\bibitem{Pich:2002xy} A.~Pich, in {\em  Phenomenology of 
large $N_C$ QCD}, ed. R.F.~Lebed (World Scientific, 2002) p. 239,
{\em arXiv:hep-ph/0205030}.

\bibitem{Kawarabayashi:1966kd} K.~Kawarabayashi and M.~Suzuki, 
{\em Phys. Rev. Lett.} {\bf 16} (1966) 255.

\bibitem{Riazuddin:1966sw} Riazuddin and Fayyazuddin, {\em Phys. Rev.} 
{\bf 147} (1966) 1071.

\bibitem{Hagiwara:2002fs} K.~Hagiwara {\em et al.}, {\em Phys. Rev.}
{\bf D66} (2002) 010001.

\bibitem{Zielinski:1984au} M.~Zielinski {\em et al.}, {\em Phys. Rev. Lett.}
{\bf 52} (1984) 1195.


\end{thebibliography}

\end{document}